\documentstyle[12pt]{article}
\newcommand{\be}{\begin{equation}}
\newcommand{\ee}{\end{equation}}

\newcommand{\r}{{\bf r}}
\input epsf
\begin{document}
\begin{center}
\large
{\bf  
A Quantum Monte Carlo Approach to the Adiabatic Connection Method \\[1cm]}
\normalsize
{\bf Maziar Nekovee and W.M.C. Foulkes \\
The Blackett Laboratory, Imperial College, \\
Prince Consort Road, London SW7 2BZ, UK\\[1cm]}
{\bf
A.J. Williamson, G. Rajagopal and R.J. Needs\\
TCM group, Cavendish Laboratory,\\
Madingley Road, Cambridge CB3 0HE, UK \\[1cm]}
\large {\bf Abstract}\\[1cm]
\normalsize
\begin{minipage}{6in}
\small
We present a new method for realizing the adiabatic connection 
approach in density functional theory, which 
is based on combining accurate variational quantum Monte Carlo calculations 
with a constrained optimization of the ground state many-body wavefunction
for different values of the Coulomb coupling constant.
We use the method to study an  electron gas in the presence of
a cosine-wave potential. For this system we present results for the 
exchange-correlation hole and exchange-correlation energy density,
and compare our findings with those from the local density approximation 
and generalized gradient approximation. 
\normalsize
\begin{enumerate}
\item  Introduction 
\item  The Adiabatic connection 
\item  Quantum Monte Carlo realization

3.1 Variational Monte Carlo

3.2 Fixed-density variance minimization
\item Cosine-wave jellium

4.1 Many-body wavefunction
\item Results and discussion
\end{enumerate}
\end{minipage}
\end{center}
\newpage
\section{Introduction}
Density functional theory (DFT) (\ref{hk}, \ref{ks}) is the main 
computational tool for
the treatment of many-body effects in solid state electronic structure 
calculations and is now widely used 
to determine ground-state properties of atoms and molecules
(\ref{parr}). 
In the Kohn-Sham formulation of DFT (\ref{ks}) the problem of finding the 
ground state energy and density of an interacting N-electron
system is tranformed into an equivalent problem involving
non-interacting electrons. The central quantity in this formulation
is the exchange-correlation energy $E_{xc}$, which is a universal
functional of the electron density $n(\r)$. 
The exchange-correlation energy functional, a complicated many-body 
object, is the big unknown of the theory and the core problem in the 
density functional approach is to find accurate approximations for $E_{xc}$.
The most frequently used approximations to date are the local 
density approximation (LDA) (\ref{ks}) and various generalized gradient 
approximations (GGA) (\ref{lm}, \ref{becke}, \ref{pd1}). 

An entirely different approach to the ground-state many-body problem
is quantum Monte Carlo (QMC) (\ref{hammond}).
QMC calculations are computationally more demanding
than density functional calculations.
However, unlike the density functional approach, in which 
the ground-state density is the basic variable, quantum Monte Carlo
methods focus on sampling the full ground-state many-body
wavefunction of the system under consideration and hence yield 
a more detailed description of many-body effects. Quantum Monte
Carlo calculations can therefore be used to investigate density 
functional theory from ``outside'' and to test
the performance of approximations to $E_{xc}$. 
In the last few years a number of 
quantum Monte Carlo investigations of DFT have been reported for 
atoms and molecules (\ref{gonze1}, \ref{gonze2}), 
model solids (\ref{godby1}, \ref{engel}) and 
silicon (\ref{randy}).
Most of these investigations focused on extracting
the exchange-correlation potential and components of exchange-correlation 
energy from accurate electron densities obtained from Monte Carlo 
calculations. Except for a very recent calculation by Hood
 {\it et al.}(\ref{randy}), 
other key quantities in DFT, namely the 
exchange-correlation hole $n_{xc}$ and the exchange-correlation energy density
$e_{xc}$, have not been investigated with Monte Carlo methods.
These quantities, however, are important in understanding the success of 
the LDA beyond its formal limits of validity,
and play a key role in constructing more accurate approximations to $E_{xc}$.
A better knowledge of these quantities is therefore crucial for a better 
understanding of the performance of the LDA  and various corrections to 
it such as GGAs, and can guide the construction of more accurate functionals.
Unlike $V_{xc}$ which can be directly obtained from the electron 
density (by inversion of the Kohn-Sham equations 
(\ref{godby1}, \ref{gonze1})) evaluating 
$e_{xc}$ and $n_{xc}$ is more demanding. These quantities are 
derived from  an adiabatic connection 
procedure in which one scales the Coulomb interaction by a factor 
$\lambda$ while keeping the density fixed at the ground state density
of the system under consideration. 
To extract $n_{xc}$ and $e_{xc}$ from Monte Carlo data one therfore needs to 
calculate not only the ground state many-body wavefunction of the
fully interacting system ($\lambda=1$),
but also the many-body wavefunction in the range $0\leq \lambda< 1$.

Within variational quantum Monte Carlo, we have developed a new scheme 
for realizing the above adiabatic connection procedure which allows
us to extract $n_{xc}$ and $\epsilon_{xc}$ from Monte Carlo data. 
Our method is based on a constrained 
optimization of the many-body wavefunction at different Coulomb 
coupling constants using the technique of variance minimization
(\ref{umrigar}, \ref{andrew1}).
In this paper we will discuss aspects of our method and 
illustrate it with a first application to an electron gas exposed to a 
cosine-wave potential.
For this  system we calculate the exchange-correlation energy, 
exchange-correlation energy density and exchange-correlation 
hole, and compare our findings with those obtained from the LDA and the 
most commonly used version of GGA (\ref{pd1}, \ref{pd2}).
\section{The Adiabatic Connection}
The idea of an adiabatic connection to determine $E_{xc}$ has been 
developed by several authors (16-18). 
Here we closely follow the review by Parr and Yang (\ref{parr}).
We consider a system of $N$ interacting 
electrons in the presence of
an external potential $V_{ex}(\r)$ and characterized by the Hamiltonian 
(atomic units are used throughout, with $ e = \hbar = m = 1$)
\be
\hat{H}= \hat{T} +\hat{V}_{ee}+\hat{V}_{ex}
\ee
with 
\be
\hat{T} = \sum_{i=1}^{N} -\frac{1}{2}\nabla^{2}_{i}
\ee
\be
\hat{V}_{ee}=
\frac{1}{2}\sum_{i=1}^{N}\sum_{j\neq i}^{N} \frac{1}{|\r_i-\r_j|}
\ee
\be
\hat{V}_{ex}=\sum_{i=1}^{N} V_{ex}(\r_i)
\ee
In the Kohn-Sham formulation of DFT
the problem of finding the ground state energy of this system is
exactly mapped onto one of finding the electron  density 
which minimizes the total energy functional
\be
E[n(\r)] = T_0[n(\r)]+E_H[n(\r)]+ 
\int\; d\r V_{ex}(\r) n(\r) + E_{xc}[n(\r)]
\ee
Here $T_0$ is the kinetic energy of a fictitious non-interacting system 
of $N$ electrons having the same electron density $n(\r)$
as the interacting system and $E_{H}[n]$ 
is the Hartree (electrostatic) energy. The exchange-correlation 
energy functional $E_{xc}[n]$ is usually defined by equation (5) and
contains all the many-body terms not considered elsewhere in (5).

An exact  expression for $E_{xc}$ is obtained 
by scaling the electron-electron interaction with a factor 
$\lambda$ and varying $\lambda$ between 1 (real system)  and 0 
(non-interacting system). 
The exchange-correlation functional $E_{xc}$ is then given by~(\ref{parr})
\be
E_{xc}[n]= \int^{1}_{0} 
d\lambda <\Psi^{\lambda}|\hat{V}_{ee}|\Psi^{\lambda}>
-E_H[n]
\ee
where $\Psi^{\lambda}$ is the anti-symmetric many-body wavefunction which 
minimizes 
\be
F^{\lambda}= <\hat{T}+\lambda \hat{V}_{ee}>
\ee
under the fixed-density constraint 
\be
<\Psi^{\lambda}|\hat{n}(\r)|\Psi^{\lambda}> =n(\r)
\ee
and $\hat{n}(\r)$ is the density operator
\be
\hat{n}(\r) =\sum_{i=1}^{N} \delta(\r-\r_i).
\ee
A minimum for $F^{\lambda}$ always exists (\ref{lieb}) and, 
except under some unusual conditions (\ref{englisch}), 
$\Psi^{\lambda}$ can be obtained from the following 
Schr\"{o}dinger equation
\be
[\hat{T}+\lambda\hat{V}_{ee} +\hat{V}^{\lambda}]\Psi^{\lambda}=
\hat{H}_{\lambda} \Psi_{\lambda} = E^{\lambda}\Psi^{\lambda}
\ee
with 
\be
\hat{V}^{\lambda} = \sum_{i=1}^{N} V^{\lambda}(\r_i).
\ee
The potential $V^{\lambda}$(\r) at point $\r$ 
is a Lagrange multiplier corresponding 
to the fixed-density constraint at that point.
As $\lambda$ varies between $0$ and $1$, $V^{\lambda}(\r)$ 
must be adjusted such that
the electron density remains fixed at $n(\r)$. At $\lambda=1$,
$V^{\lambda}$ coincides with the actual external potential 
$V_{ex}(\r)$  while at $\lambda=0$, it coincides with the 
Kohn-Sham effective potential,
\be
V^{\lambda=0}(\r) = V_{eff}(\r)= V_{ex}(\r)+V_H(\r)+V_{xc}(\r) 
\ee
where $V_{H}$ is the Hartree (electrostatic) potential
\be
V_{H}(\r) = \int d \r' \frac{n(\r')}{|\r-\r'|}
\ee
and 
\be
V_{xc}(\r) = \frac{\delta E_{xc}[n]}{\delta n(\r)}
\ee  
is the Kohn-Sham exchange-correlation potential. 
Note also that $\Psi^{\lambda=0}$ corresponds to the Slater determinant 
of the exact Kohn-Sham orbitals corresponding to the density $n(\r)$.

The adiabatic expression (6) allows us to obtain several useful 
decompositions of $E_{xc}$. Inserting (3)  in  (6) gives
\be
E_{xc}[n({\bf r})]  =  
\frac{1}{2}\int d\r \int d\r'\; 
\frac{n(\r) n_{xc}(\r,\r')}{|\r -\r'|}
\ee
where $n_{xc}$ is the density-functional exchange-correlation 
hole defined by (\ref{parr}) 
\be
\bar{n}(\r,\r') = n(\r)n(\r')+n(\r)n_{xc}(\r,\r')
\ee
Here $\bar{n}(\r,\r')$ is the diagonal part of 
the two-particle density matrix averaged over $\lambda$,
\be
\bar{n}(\r,\r')=  
\int_{0}^{1} d\lambda\; n^{\lambda}(\r,\r')
\ee
and
\be
n^{\lambda}(\r,\r') =
 <\Psi^{\lambda}| 
\sum_{i=1}^{N} \sum_{j\neq i}^{N} \delta(\r-\r_i)\delta(\r'-\r_j)
|\Psi^{\lambda}>.
\ee
Integrating (15)  over $\r'$ yields 
\be
E_{xc}[n({\bf r})]  =  
\int d\r\; e_{xc}(n[\r],\r)
\ee
where $e_{xc}$ is the exchange-correlation energy density 
derived from the adiabatic connection procedure
\be
e_{xc}([n(\r)],\r)= \int_{0}^{1} d\lambda\; e^{\lambda}_{xc}([n(\r)],\r)
\ee
with $e^{\lambda}_{xc}$ given by
\be
e^{\lambda}_{xc}([n(\r)],\r)=
<\Psi_{\lambda}| \frac{1}{2}\sum_{i=1}^{N}\sum_{j\neq i}^{N}
\frac{\delta(\r-\r_i)}{|\r-\r_j|}| \Psi_{\lambda}> - \frac{1}{2} 
\int d \r' \frac{n(\r)n(\r')}{|\r-\r'|}
\ee

For further reference we note that $n_{xc}^{\lambda=0}$
corresponds to the {\em density functional} exchange hole $n_{x}$.
The corresponding excahange energy density is $e_x = e_{xc}^{\lambda=0}$
and the correlation energy density is given by $e_c=e_{xc}-e_x$.
Note, however, that the excahnge-correlation energy density, and hence 
its exchange and correlation components, are not uniquely defined
quantities since we can always add to $e_{xc}$ any function which 
integartes to zero without affecting the exchange-correlation energy.
Our definition of these quantities emerges in a natural way from the 
adiabatic connection. 
An alternative definition of the correlation energy density
has been suggested by Baerends and Gritsenko (\ref{barends}) and 
by Huang and Umrigar (\ref {huang}).

\section{Quantum Monte Carlo realization}
Given an interacting many-body system with ground-state density 
$n(\r)$, the main ingredient for evaluating 
$n_{xc}$ and $e_{xc}$ is the 
many-body wavefunction $\Psi^{\lambda}$ for a number of systems
corresponding to different values of the coupling constant $\lambda$
satisfying the fixed-density constraint.
In this section we describe our variational  quantum Monte 
Carlo algorithm for obtaining $\Psi_{\lambda}$.
\subsection{Variational Monte Carlo}
In variational Monte Carlo calculations (\ref{hammond}) 
one starts off with an explicit 
parameterized {\it Ansatz} for the ground-state many-body
wavefunction of the system under consideration. The total energy 
of the system is then calculated as the expectation value of 
the Hamiltonian $\hat{H}$ with respect to the variational wavefunction
$\Psi_{T}$.
Monte Carlo integration is used to perform  the multi-dimensional 
integrals required for evaluating this expectation value and the 
variational parameters in $\Psi_{T}$ are adjusted until an optimized 
wavefunction is obtained. The 
state-of-the-art method for performing the optimization procedure 
is the variance minimization scheme (\ref{umrigar}, \ref{andrew1}). 
In this scheme one minimizes 
the variance of the local energy $\hat{H}\Psi_{T}/\Psi_{T}$ (rather
than expectation value of $\hat{H}$) with respect to variational 
parameters over a set of particle configurations. 
The use of energy optimized wavefunctions may 
give unsatisfactory results when quantities other than the energy are 
evaluated, while minimization 
of the variance tends to give a better fit for the wavefunction 
as a whole, so that satisfactory results are obtained for 
a range of quantities including both energy {\em } and electron
density.  The electron density plays a central role in the adiabatic
connection procedure making variance minimization a more suitable 
choice for optimizing $\Psi^{\lambda}$.
\subsection{Fixed-density variance minimization}
We consider an N-electron system having ground-state density 
$n(\r)$. At a given coupling constant 
$\lambda$ the corresponding many-body wavefunction 
$\Psi^{\lambda}$ satisfies equation (10). 
Therefore, at an arbitrary point ${\bf R}=(\r_1,\r_2,\ldots \r_N)$ 
in the $3N$ dimensional configuration space of electron coordinates, 
we have   
\be
\frac{H^{\lambda}\Psi^{\lambda}({\bf R})}
{\Psi^{\lambda}({\bf R})} - E^{\lambda} \equiv 0.
\ee
In conventional variance minimization calculations (\ref{andrew1}) 
(i.e. the unconstrained $\lambda=1$ case),
the above property is used to find an overall fit to $\Psi$
(we drop the $\lambda$ superscript for simplicity).
The procedure is to 
determine the parameters $\{\alpha\}$ in the trial function 
$\Psi_{T}({\bf R},\{\alpha\})$ 
by minimizing the variance of local energy $\sigma^2$
\be
\sigma^{2} = \int d{\bf R}\;  
\left[\frac {H^{\lambda}\Psi_T({\bf R})}
{\Psi^{\lambda}_T({\bf R})}  - E[\Psi_T]\right]^{2}
|\Psi_{T} ({\bf R})|^2
\ee
where $E[\psi_T]$ is the expectation value of the Hamiltonian.

The above unconstrained optimization cannot be directly 
applied at intermediate values of $\lambda$ for which the Hamiltonian 
contains the unknown potential $V_{\lambda}$. 
We found, however, that a simultaneous determination of 
$\Psi_{\lambda}$ and $V_{\lambda}$ can be achieved by performing
the following constrained optimization. 
We assume that the trial many-body 
wavefunction $\Psi^{\lambda}_{T}$ results in the electron density 
$n^{\lambda}(\r)$ and expand both $n^{\lambda}(\r)$ and 
the ground state density $n(\r)$ in a complete and orthonormal set of 
basis functions $\{f_s\}$ 
\be
n(\r) = \sum_{s=1}^{N_d} n_s f_s(\r)
\ee
\be
n^{\lambda}(\r) = \sum_{s=1}^{N_d} n^{\lambda}_sf_s(\r)
\ee
where $N_d$ is a cut-off chosen such that the above expansions converge 
to $n(\r)$ and $n_{\lambda}(\r)$ within a specified accuracy. 
Subsequently, we define the modified penalty function $\mu^2$
\be
\mu^{2} = \sigma^{2} +W  
\sum_{s=1}^{N_d}\left[n_s-n_{s}^{\lambda}\right]^2  
\ee
where $W$ is a weight factor the magnitude of which determines the 
emphasis laid on the fixed-density constraint.
The above penalty function reaches its lower bound (of zero) 
if and only if $\Psi^{\lambda}$ is the 
exact many-body wavefunction satisfying the fixed density 
constraint (within the accuracy set by $N_d$) 
and $V^{\lambda}$  is the corresponding 
exact potential. Hence minimization of 
$\mu^{2}$ will, in principle, result in the {\em simultaneous}
determination of $\Psi^{\lambda}$ and $V^{\lambda}$.
In practice, however, our constrained 
search is restricted to a sub-space of many-body wavefunctions and 
minimization of $\mu^{2}$ yields an optimal fit to $\Psi^{\lambda}$ 
and a corresponding optimal fit to $V^{\lambda}$, the deviations
of which from the exact $V^{\lambda}$ reflect the errors in 
the many-body wavefunction.

Our numerical implementation of the above scheme works as follows.
We start off with an initial guesses  
$\Psi^{\lambda}_{0}$ for the many-body wavefunction 
and  a corresponding guess for $V^{\lambda}$.  
A fixed number $N_c$ of statistically independent 
configurations ${\bf R}_i$ are then sampled from $|\Psi_{0}^{\lambda}|^{2}$
and the Monte Carlo estimator of $\mu^{2}$ over these configurations 
is evaluated
\be
\mu^{2} = \sum_{i=1}^{N} \left(E_L({\bf R}_i)-<E_L>\right)^{2} 
\left[\frac {\omega_i}{\sum_{j=1}^{N_c}\omega_j}\right] 
+W\sum_{s=1}^{N_d}[n_s-n_{s}^{\lambda}]^{2}
\ee
with 
\be
E_{L}({\bf R}_i) = \frac{H\Psi^{\lambda}_T({\bf R}_i)}
{\Psi_T^{\lambda}({\bf R}_i)}
\ee
\be
\omega_i = \left|\frac{ \Psi_{T}^{\lambda}}
{\Psi_{0}^{\lambda}}\right|^{2}
\ee
$<E_{L}>$ the average energy
\be
<E_L> = \sum_{i=1}^{N_c} E_L({\bf R}_i)
\left[\frac{\omega_i} {\sum_{j=1}^{N_c}\omega_j} \right] 
\ee
The expansion coefficients of the electron density, $n_s^{\lambda}$,
are evaluated from
\be
n_{s}^{\lambda} = 
\sum_{i=1}^{N_c} \sum_{k=1}^{N} f^*(\r_{k}^{i})
\left[\frac {\omega_i}{\sum_{j=1}^{N_c}\omega_j}
\right]
\ee 
where $\r_{k}^{i}$ denotes the coordinates of the electron $k$ belonging 
to configuration $i$. 
Finally, we vary the parameters in  
$\Psi_{T}^{\lambda}$ and $V^{\lambda}$, using a standard NAG routine for
optimization, until $\mu^{2}$ is minimized.
We found that setting $W$ equal to the number of configurations results in a 
satisfactory minimization of both the variance in energy and 
the error in electron density.

Following (\ref{andrew1}) we set the reweighting factors $\omega_i$ 
in equations (27) and (30) equal to unity in order to 
avoid a numerical instability in the variance 
minimization procedure which occurs for systems with a large number 
of electrons (these factors, however, are included in calculating 
the expansion coefficients of the electron density).
The above fixed-density variance minimization is then 
repeated several times until the procedure converges. 

\section{Cosine-wave jellium}
We performed adiabatic connection calculations for the 
inhomogeneous spin-unpolarized electron gas with average electron density
$n_0=3/(4\pi r_s^3)$ corresponding to $r_s=2$.
In the QMC simulations we model this system 
by a finite system of $N=64$ electrons satisfying periodic boundary conditions
in a FCC simulation cell. Density modulations can be 
induced by applying an external potential of the form
$V_{{\bf q}} \cos({\bf q}.\r)$
where, because of periodic boundary conditions, 
${\bf q}$ is restricted to be a reciprocal lattice vector of the 
simulation cell.
Alternatively, we can fix the ground-state electron density 
{\em a priori} and use our fixed-density variance minimization method 
to obtain the corresponding many-body wavefunction at a given Coulomb 
coupling constant which produces the specified density.
In the calculations reported here we chose this second option, 
with the ``target'' electron density for the system generated in the 
following way. We expose 
the non-interacting electrons (i.e. the $\lambda=0$ system) 
to the potential $V(\r)$ 
\be
V(\r) = V_{{\bf q}} \cos({\bf q}.\r)
\ee
with $V_{{\bf q}}=2.084\epsilon_{F}^0$ and ${\bf q}=2{\bf B_3}$ . Here
$\epsilon_F^0$ is the Fermi energy 
of the unperturbed electron gas, ${\bf B_3}$ is a primitive vector 
of the reciprocal (simulation) cell with $|2{\bf B_3}|=1.11k_{F}^0$, 
and $k_{F}^0$  is the Fermi wavevector.  
We then solve the following self-consistent 
single-particle Schr\"{o}dinger equations
\be
[-\frac{1}{2}\nabla^{2}+V_{eff}]\phi_i = \epsilon_i \phi_i
\ee
with  
\be
V_{eff}(\r) = V(\r)+V_{H}(\r)+V_{xc}^{LDA}(\r)
\ee
to obtain the electron density
\be
n(\r) = 2\sum_{i=1}^{N/2} |\phi_i(\r)|^{2}
\ee
We {\em define} this density to be the exact ground-state 
density of our interacting system. In this way, the single-particle
orbitals $\phi_i$ are by construction the exact Kohn-Sham orbitals
and their Slater determinant corresponds exactly to the many-body 
wavefunction at $\lambda=0$. Having obtained this non-interacting 
$v-$representable density we then 
perform fixed-density variance minimization to produce variational 
many-body wavefunctions at non-zero Coulomb coupling constants
(including the ground-state many-body wavefunction at $\lambda=1$)
which reproduce this density and (variationally) satisfy the
Schr\"{o}dinger equation (10). Once the $\Psi_{\lambda}$s are obtained,
we use the  Monte Carlo Metropolis algorithm to evaluate the 
required expectation values and perform a numerical coupling 
constant integration using Gaussian quadrature.
\subsection{Many-body wavefunction}
The quality of a variational quantum Monte Carlo calculation is determined 
by the choice of the many-body wavefunction.
The many-body wavefunction we use is of the parameterized 
Slater-Jastrow type which has been shown to yield accurate results
both for the homogeneous electron gas and for solid silicon
(\ref{andrew1}) 
(In the case of silicon, for example, $85\%$ of the fixed-node diffusion 
Monte Carlo correlation energy is recovered).
At a given coupling $\lambda$, $\Psi^{\lambda}$ is 
written as
 \be
\Psi ^{\lambda} = 
D^{\uparrow }D^{\downarrow }\exp \left[ -\sum_{i>j}u^{\lambda}_{\sigma_i,
\sigma_j}
(r_{ij})+\sum_i\chi^{\lambda} ({\bf r}_i)\right] 
\end{equation}
where $r_{ij}=|\r_i-\r_j|$ and
$D^{\uparrow }$ and $D^{\downarrow }$ are Slater determinants of
spin-up and spin-down Kohn-Sham orbitals respectively.
$u^{\lambda}_{\sigma_i,\sigma_j}$ is the two-body term correlating 
the motion of pairs of electrons and
$\sigma_{i}$ denotes the spin of electron $i$. 
Finally, $\chi^{\lambda}$ is a one-body 
function which is absent in the homogeneous 
electron gas but is crucial for a satisfactory description 
of systems with inhomogeneity. 
Both $u^{\lambda}$ and $\chi^{\lambda}$
contain variational parameters. We write $u^{\lambda}$ as
(\ref{andrew1})
\be
u^{\lambda}(r)= u_{0}^{\lambda}(r)+f^{\lambda}(r),
\ee
where $u_{0}^{\lambda}$ is a fixed function and $f^{\lambda}$ is given by
\begin{equation}
\begin{array}{llll}
f^{\lambda}(r) & = & B^{\lambda}
(\frac{L_{WS}}{2}+r)(L_{WS}-r)^2+r^2(L_{WS}-r)^2\sum_{l=0}^M\alpha
_l^{\lambda}T_l(\overline{r})\;\;\;\; & 0\leq r\leq L_{WS} \\ 
& = & 0 & r>L_{WS}
\end{array}
\end{equation}
where $B^{\lambda}$ and $\alpha _{l}^{\lambda}$ are
variational coefficients, $T_l$ is the $l$th
Chebyshev polynomial, and
\begin{equation}
\overline{r}=\frac{2r-L_{WS}}{L_{WS}}.
\end{equation}
In the last two equations $L_{WS}$ is the 
radius of the sphere touching the Wigner-Seitz cell of the simulation 
cell.

The fixed part of $u^{\lambda}$ at full coupling constant
$\lambda=1$ is the short-ranged Yukawa form (\ref{andrew1}) 
\begin{equation}
u_0^{1}(r)=\frac{A^{1}}{r}\left( 1-\exp (-\frac {r}{F^{1}})
\right) \exp \left( -\frac{r^2}{L_0^2%
}\right) ,
\end{equation}
where $A^1$ is fixed by the plasma frequency of the unperturbed electron gas
\be
A^1= \frac{1}{\omega_p^0}
\ee
and $F^1$ is fixed by imposing the cusp condition (\ref{hammond}) leading to
$F^1_{\sigma_i,\sigma_j}=\sqrt(2A^1)$ for parallel spins and 
$F^1_{\sigma_i,\sigma_j}=\sqrt(A^1)$ for anti-parallel spins.  
$L_0$ is a cut-off parameter 
chosen so that $u_0(L_{WS})$ is effectively zero and is set
equal to $0.25L_{WS}$ in the present calculations.
In the case of the unperturbed electron gas, scaling arguments (\ref{levy3}) 
applied to the Hamiltonian (10) result in the following relation 
for the exact many-body wavefunction at coupling constant~$\lambda$ 
\be
\Psi^{\lambda}_{r_s}(\r_1,\r_2,\ldots \r_n) = 
C^{\lambda}\Psi^{\lambda=1}_{r'_s}
(\lambda \r_1,\lambda \r_2,\ldots \lambda \r_n)
\ee
where $\Psi^{\lambda=1}_{r'_s}$ is the ground-state wavefunction 
of a homogeneous electron gas with the density parameter $r'_s=\lambda r_s$
and $C^{\lambda}$ is a normalization constant.
For the unperturbed electron gas ($\chi \equiv 0$) 
imposing condition (42) on the fixed-part of our Slater-Jastrow 
wavefunction yields
\begin{equation}
u_0^{\lambda}(r)=\frac{A^{\lambda}}{r}\left( 1-\exp (-\frac {r}{F^{\lambda}
})
\right) \exp \left( -\frac{r^2}{L_0^2%
}\right) ,
\end{equation}
where $A^{\lambda}= \lambda^{1/2}A^1$, $F^{\lambda}= \lambda^{-1/4}F^1$.
We note that with the above choice for $A^{\lambda}$ and $F^{\lambda}$
the $\lambda-$dependent cusp conditions are automatically 
satisfied. 
The electron density is modulated only in the ${\bf B_3}$ direction
and hence both the one-body part of the Jastrow factor and 
$V^{\lambda}$ can be expanded as
\be
\chi^{\lambda}(\r)=\sum_{m=1}^{M}\chi^{\lambda}(m{\bf B_3}) 
\cos(m{\bf B_3}.\r)
\ee
\be
V^{\lambda}(\r)=\sum_{m=1}^{M}V^{\lambda}(m{\bf B_3}) 
\cos(m{\bf B}_3.\r)
\ee
The electron density is expanded in a similar way (with the inclusion of 
the $m=0$ term). 
We use $7$ Fourier coefficients in the expansion of electron density,
$6$ Fourier coefficients in the expansions of $\chi_{\lambda}$ 
and $V_{\lambda}$ (only the first four coefficients turned 
out to be significantly different from zero),  
and $8$ coefficients (for each of the spin-parallel and spin-antiparallel 
cases) in the two-body term.

\section{Results and discussion}
\begin{figure}[t]
\epsfxsize=12 truecm
\centerline{\epsffile{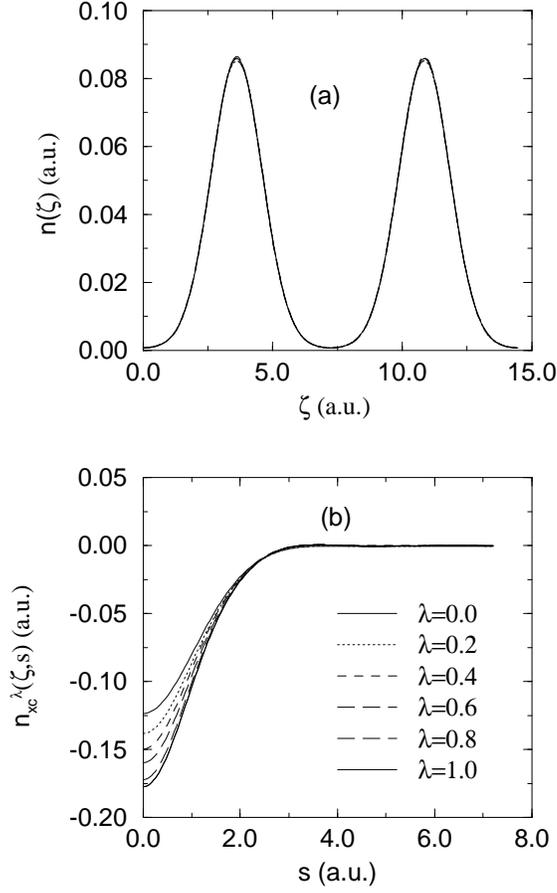}}
\caption{{ \small (a) Electron density for different values of $\lambda$
plotted along $\zeta$ 
(the direction of inhomogeneity). (b) 
The $\lambda-$dependent 
spherically averaged exchange-correlation hole for an electron 
sitting at $\zeta=10.85$ a.u.}}
\end{figure}

We performed adiabatic connection calculations for cosine-wave jellium
using six values of $\lambda$: $0,0.2,0.4,0.6,0.8,1$.
The many-body wavefunctions for $\lambda>0$ were optimized 
by fixed-density variance minimization using 
$10000$ independent $N-$electron configurations at each 
$\lambda$. These configurations were regenerated several times.
The weight factor in expression (27) was set equal to 
$10000$ in order to obtain a satisfactory minimization of both the variance 
in energy and the error in electron density.
Once $\Psi_{\lambda}$ was optimized, quantities of interest were accumulated 
with the Metropolis Monte Carlo algorithm using $500000$ statistically
uncorrelated configurations.

We found that our method results in electron densities $n^{\lambda}(\r)$
which deviate from the reference density by less than $1\%$.
This is shown in figure 1(a) where the density is plotted 
as a function of $\lambda$ along a line parallel to the
direction in which the external potential varies (we call this the $\zeta$
direction). While the density is fixed, 
all other physical quantities vary smoothly and monotically with $\lambda$. 
As an example we consider the  
spherically-averaged exchange-correlation hole
\be
\tilde{n}^{\lambda}_{xc}(\r,s)= \frac{1}{4 \pi} \int_{\Omega} d\r'
n^{\lambda}_{xc}(\r,\r'), \; \; \; \; \Omega:|\r -\r'|=s
\ee
as a function of $\lambda$.
In figure 1(b) this quantity is shown around an electron sitting 
at one of the maxima of the electron density ($\zeta=10.85$ a.u.). 
The $\lambda=0$ curve corresponds to the spherically-averaged exchange hole.
The exchange hole is relatively shallow and negative everywhere.
As the interaction is switched on, the hole around the electron becomes
gradually deeper. 
The spherically-averaged hole obeys the sum-rule (\ref{parr})
\be
4\pi \int s^2 \tilde{n}^{\lambda}_{x}(\r,s) =-1
\ee
and the deepening of the hole for $\lambda>0$ is compensated by 
the fact that the hole becomes slightly positive 
far away from the electron.
Note that the hole does not ``narrow'' as it deepens, but actually broadens.
In evaluating $\tilde{\rho}_{xc}^{\lambda}$ we expanded 
the exchange-correlation hole in a double Fourier series, sampled the 
corresponding expansion coefficients
and subsequently performed the spherical averaging.
Our calculated $\rho_{xc}^{\lambda}$ does not satisfy the 
Kimball cusp-condition (\ref{kimbal}, ~\ref{hole}) because of the finite 
number of plane-waves in its Fourier expansion. As a result  
$\tilde{\rho}_{xc}^{\lambda}$ has zero slope at $s=0$.
We note, however, that this deficiency does not affect $E_{xc}$
and $e_{xc}^{\lambda}$ because these quantities 
are evaluated directly from equations (6) and (21).
 
\begin{figure}[t]
\epsfxsize=13 truecm
\centerline{\epsffile{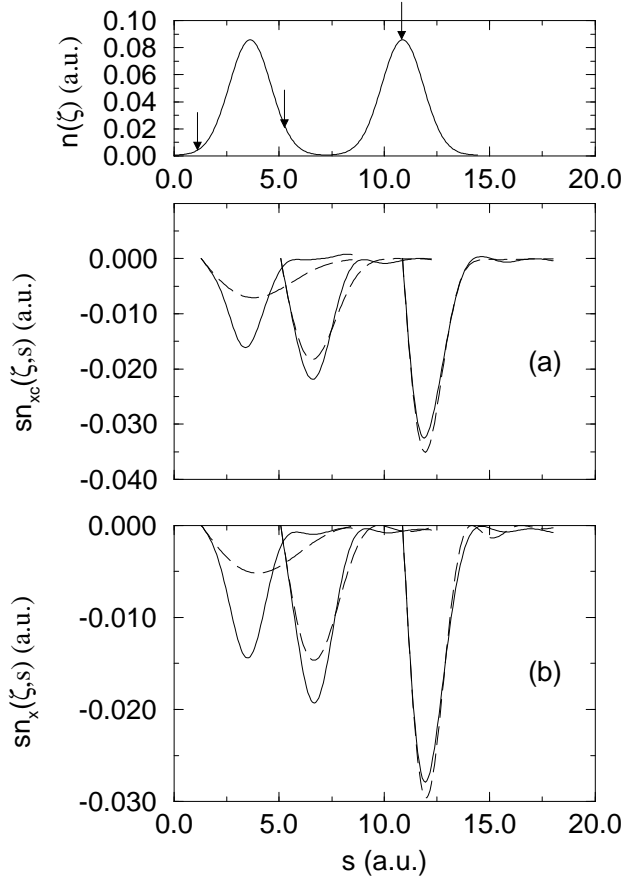}}
\caption{ {\small (a) VMC (solid lines) and
LDA (dashed lines) $s \tilde{n}_{xc}(\r,s)$
plotted for an electron moving along $\zeta$. 
Arrows on the electron density
(plotted on top), mark the position of the electron.
(b) Exact $s \tilde{n}_{x}(\r,s)$ plotted in the same direction
and at the same points as in (a) (solid lines) 
and the corresponding LDA approximation (dashed lines).}}
\end{figure}

Before discussing our findings for this system,
we would like to pause and give a short outline of 
the errors present in our simulations. 
First of all, the small ($<1\%$) deviations of the electron density 
at different $\lambda$ from 
the reference density $n(\r)$ will induce errors in the 
adiabatically calculated quantities such as the correlation energy density. 
By recalculating the exchange energy density with a density which 
deviates from the reference density by $1\%$ and extrapolating
the resulting deviation $e_x[n(\r)]-e_x[n(\r)+\delta n(\r)]$
to the correlation energy density, we estimate the 
errors in $e_c$ due to these density deviations to be also $\sim 1\%$. 

Further, there are two other kind of errors in our calculations: 
(i) statistical errors; and
(ii) finite size errors (i.e. those caused by the fact that we are 
using a finite number of electrons to model a supposedly infinite system).
With $500000$ configurations used in sampling
all physical quantities, we found statistical errors to be unimportant, 
except for the exchange-correlation hole. 
By evaluating the exchange hole both directly and by Monte Carlo sampling,
and assuming that the errors in $n_{xc}$ for $\lambda\neq0$ are 
similar,
we estimate the statistical error in this quantity to be less than $7\%$. 
Another source of errors is finite size effects. 
These errors occur because a 
finite simulation cell is
used to model an infinite system, with the Coulomb interaction energy
evaluated using the Ewald formula (\ref{matthew}).
The use of a finite simulation cell with periodic boundary conditions 
affects the wavefunction, of course, and the use of Ewald interaction 
also produces a Coulomb finite size error in the interaction energy
(\ref{matthew}, \ref{andrew2}).  
We found the effect of the finite cell on the exchange-correlation hole
to be unimportant, except for the asymptotic behavior of this quantity 
which cannot be correctly described with the present system size.
Coulomb finite size effects do not significantly affect 
the many-body wavefunctions, and hence their effect on quantities 
such as the electron density and the exchange-correlation hole is negligible.
The exchange-correlation energy density, however, is 
directly affected by Coulomb finite size effects 
since in evaluating $e_{xc}^{\lambda}$ 
the $1/|\r-\r'|$ Coulomb interaction in (21) 
is replaced by the periodic Ewald interaction.
By calculating the exchange energy density of the homogeneous 
electron gas using our finite simulation cell  and comparing it with the 
exact result
(\ref{fulde})
\be
e_x = -\frac{0.45805}{r_s} n_0 
\ee
we estimate the total finite-size error in 
$e_{x}$ to be of the order of $-2\times 10^{-4}$ a.u; 
the errors in the correlation energy density $e_c$ is expected
to be somewhat smaller.
\begin{figure}[t]
\epsfxsize=12 truecm
\centerline{\epsffile{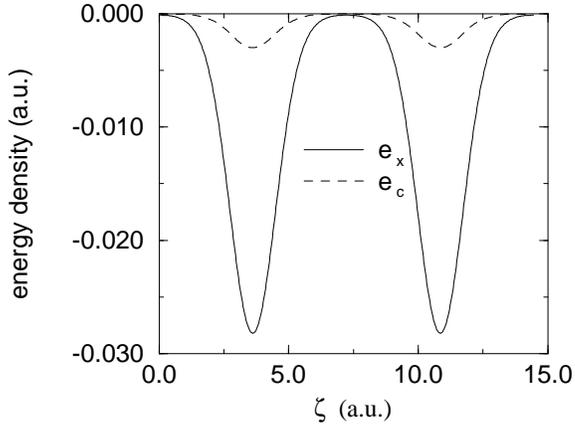}}
\caption{{\small Exchange (solid line) and correlation (dashed line) 
contributions to $e_{xc}$ plotted along $\zeta$.}}  
\end{figure}

We now turn to our results for $n_{xc}$ and $e_{xc}$.
The spherically averaged exchange-correlation hole, 
$\tilde{n}_{xc}^{\lambda}(\r,s)$, obtained from our adiabatic calculations 
is shown in figure 2(a) together with the LDA approximation~(\ref{fulde}) 
to this quantity 
\be
n_{xc}^{LDA}(\r,s) = n(\r)(\bar{g}^{hom}(n(\r),s)-1) 
\ee
where $\bar{g}^{hom}$ is the $\lambda$-averaged pair-correlation 
function of a homogeneous electron gas with density $n(\r)$
(we use the Perdew-Wang parameterization of $\bar{g}$ (\ref{hole})).
In this figure we plot $n_{xc}(\r,s)$ for an electron 
moving along $\zeta$, and hence fully experiencing the strong variations 
in electron density. 
The hole is shown multiplied by $s$ so that the area under 
each curve is directly proportional to the exchange-correlation
energy per electron $e_{xc}/n$.
At $\zeta=1.30$ a.u. the electron density is very low 
and $n_{xc}$ is shallow. As electron moves to higher densities
($\zeta=5.6$ and $\zeta=10.85$ a.u.) the hole 
becomes deeper and its asymptotic tail less pronounced. 
Unlike the LDA hole which depend only on the local density 
$n(\r)$ and is dug out of a homogeneous electron gas of that density, 
the VMC hole depends on the density everywhere in the 
vicinity of $\r$.
At $\zeta=1.30$ a.u, where the electron density is very low, the 
LDA ``probes'' only this density and for this reason the LDA 
exchange-correlation 
hole is very different from the VMC hole, even close to the electron. 
As the electron moves to the high density region, the LDA description 
becomes more satisfactory and at $\zeta=10.85$ a.u., 
where the electron density has a maximum, the agreement between the LDA 
and the VMC hole is rather good.
In figure 2(b) we compare the exact exchange hole (obtained from our exact 
Kohn-Sham orbitals) for the same electron positions 
with the LDA hole given by (\ref{fulde})
\be
\tilde{n}_{x}^{LDA}(\r,s) = -\frac{9}{2}n(\r)
\left[\frac{j_{1}(k_F(\r)s)}{k_F(\r)s}\right]^{2},
\ee
where $k_F(\r)=(3 \pi^2 n(\r))^{1/3}$ is the local Fermi wavevector
and $j_{1}$ the first order spherical Bessel function.
Once again, the LDA description is unsatisfactory 
at low densities but improves as we move to the high density region.
\begin{figure}[h]
\vspace{2cm}
\epsfxsize=12 truecm
\centerline{\epsffile{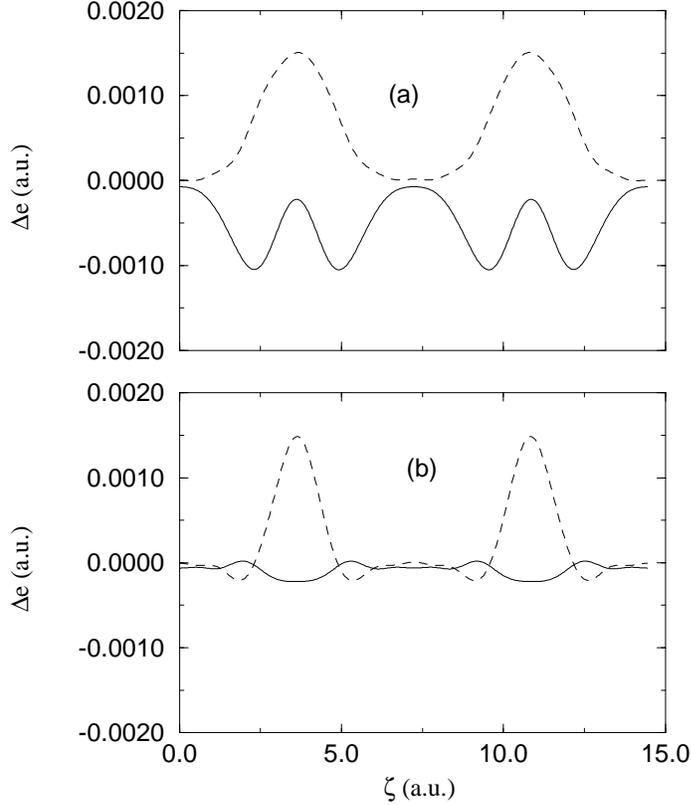}}
\caption{{\small (a) The differences $e_{x}^{VMC}-e_{x}^{LDA}$
(solid line) and $e_{c}^{VMC}-e_{c}^{LDA}$ (dashed line)
plotted along $\zeta$.
(b) The same as (a) but for
$e_{x}^{VMC}-e_{x}^{GGA}$ and $e_{c}^{VMC}-e_{c}^{GGA}$. }}
\end{figure}

Next we consider exchange-correlation energy densities.
In figure 3 the exchange and correlation contributions 
to this quantity are shown.
The differences $e_{x}^{VMC}-e_{x}^{LDA}$ and $e_{c}^{VMC}-
e_c^{LDA}$ are shown in figure 4 (a). The difference in 
$e_{c}$ follows the variations in electron density and is largest 
at points where $n(\r)$ has a maximum. The differences 
in $e_{x}$ shows a more complicated structure and 
$e_x^{VMC}< e_x^{LDA}$ everywhere in the system.
This result is in line  with the well-known fact that LDA almost
always underestimates the exchange energy of an inhomogeneous 
system.
However, because of the finite size errors, we expect 
the true $e_x$ to be slightly ($\sim 2\times 10^-4$ a.u.) less negative 
than $e_x^{VMC}$ so that $e_x<e_x^{LDA}$ must holds in most points 
of the structure but not necessarily everywhere.

The GGAs  for exchange and correlation of a spin-unpolarized system 
are written as (\ref{pd2}) 
\be
E_x = \int d\r \; n(\r)\epsilon_{x}^{unif}(n(\r))F_x(s)
\ee
\be
E_c = \int d\r \; n(\r)[\epsilon_c^{unif}(n(\r))+H(n(\r),t)]
\ee
In the above equations $\epsilon_x^{unif}$ and 
$\epsilon_c^{unif}$ are the exchange
and correlation energies per particle of a uniform electron gas with 
density $n(\r)$, $t=|\nabla n|/2k_s(\r)$, $ s= |\nabla n|/2k_F(\r)$
with $k_s$ the local Thomas-Fermi wavevector 
$k_s(\r)=\sqrt{4k_F(\r)/\pi}$.
In analogy with (19) one may define the GGA $e_{x}$ and $e_{c}$
as
\be
e_{x}^{GGA}(\r) = n(\r)\epsilon_{x}^{unif}(n(\r))F_x(s)
\ee
and 
\be
e_{c}^{GGA}(\r) = n(\r)[\epsilon_c^{unif}(n(\r))+H(n(\r),t)]
\ee
We note that the above quantities do not directly
correspond to the physical $e_x$ and $e_{c}$ as defined by 
equation (21) (i.e. via the exchange-correlation hole).
Nevertheless, they present pointwise corrections to the LDA 
exchange and correlation energy densities and for this reason
we found it interesting to compare them with our VMC results.
In figure 4(b) we show $e_{x}^{VMC}-e_{x}^{GGA}$ and $e_{c}^{VMC}-
e_c^{GGA}$. The GGA results were obtained from the ground state 
density $n(\r)$ using the Perdew-Burke-Ernzerhof scheme (\ref{pd2}), 
which we found  to give results slightly different from PW91 (\ref{pd1}).
Note that the difference
between the VMC and the GGA exchange energy densities 
is significantly smaller than that between
the VMC and the LDA exchange energy densities, 
indicating that the GGA improves upon the LDA in describing 
this quantity. As for $e_c$, the differences are of 
the same order as for the LDA, although the shape is different.
It is interesting that the LDA errors in $e_x$ and $e_c$ partially
cancel each other, even on a local scale, but that these cancellations 
do not occur for the GGA. In summary, the GGA seems to do a good job 
in improving the LDA description of the exchange energy density 
but is less successful in the case of the correlation energy density. 
The resulting exchange correlation energy 
per electron  obtained from VMC, LDA and GGA are 
$E_{xc}^{VMC}/N = -0.328 \pm 0.009$, $E_{xc}^{LDA}/N= -0.3296$, 
$E_{xc}^{GGA}/N=-0.3347$ a.u.
\section*{Acknowledgments}
We thank Randy Hood and Mei-Yin Chou for helpful discussions.
Part of this work has been supported by the Human Capital and Mobility 
Program through contract No. 
CHRX-CT94-0462.

\section*{References}
\begin{list}
{(\arabic{enumi})}{\usecounter{enumi}}
\item\label{hk} P. Hohenberg and W. Kohn, Phys. Rev. {\bf 136}, B864 (1964).
\item\label{ks} W. Kohn and L.J. Sham, Phys. Rev. {\bf 140}, A1133 (1965).
\item\label{parr} R.G. Parr and W. Yang,
{\em Density Functional Theory of Atoms and Molecules} 
(Oxford University Press, New York, 1989). 
\item\label{lm} 
D.C. Langreth and M.J. Mehl, Phys. Rev. B {\bf 28},
1809 (1983).
\item\label{becke} A.D. Becke, Phys. Rev. A {\bf 38}, 3098 (1988).
\item\label{pd1} J.P. Perdew {\it et al.}, Phys. Rev. B {\bf 46}, 6671 (1992);
{\bf 48}, 4978 (1993) (E).
\item\label{hammond} B.L. Hammond, W.A. Lester, Jr. and R.J. Reynolds,
{\em Monte Carlo Methods in Ab Initio Quantum Chemistry} 
(World Scientific, Singapore, 1994).
\item\label{gonze1} C.J. Umrigar and X. Gonze in {\em High Performance 
Computing and its Applications to the Physical Science}, edited by
D.A. Browne {\it et al.} (World Scientific, Singapore, 1993). 
\item\label{gonze2} C. Filippi, C.J. Umrigar  and X. Gonze,
Phys. Rev. A {\bf 54}, 4810 (1996).
\item\label{godby1} W. Knorr and R.W. Godby, Phys. Rev. Lett. {\bf 68}, 639 
(1992).
\item\label{engel} G.E. Engel, Y. Kwon and R.M. Martin, 
Phys Rev. B {\bf 51} (1995).
\item\label{randy} R.Q. Hood {\it et al.} Phys. Rev. Lett. {\bf 78},
3350 (1997).
\item\label{umrigar}
C.J. Umrigar, K.G. Wilson and J.W. Wilkins, Phys. Rev. Lett. {\bf 60},
1719 (1988).
\item\label{andrew1} A.J. Williamson {\it et al.},
Phys. Rev. B {\bf 53}, 9640 (1996).
\item\label{pd2} J.P. Perdew, K. Burke and M. Ernzerhof,
Phys. Rev. Lett. {\bf 77}, 3865 (1996).
\item\label{jones} J. Harris and R.O. Jones, J. Phys. F {\bf 4}, 1170 (1974).
\item\label{langr} D.C. Langreth and J.P. Perdew, Phys. Rev. B {\bf 21},
5469 (1980).
\item\label{lund} O. Gunnarson and B.I. Lundqvist, Phys. Rev. B {\bf 13},
4274 (1976). 
\item\label{levy1} M. Levy in {\em Density Functional Theory}, edited 
by J. Keller and J.L. Gasquez (Springer, New york, 1983).
\item\label{lieb} E.H. Lieb, in {\em Physics as Natural Philosophy},
edited by A. Shimony, H. Feshbach (MIT Press, Cambridge, Mas, 1982).
\item\label{englisch} H. Englisch and R. Englisch, Physica (Utrecht)
{\bf 21A}, 253 (1983).
\item\label{barends} E.J. Baerends and O.V. Gritsenko J. Phys. Chem., 
{\bf 101}, 5383 (1997). 
\item\label{huang} 
Chien-Juang  Huang and C.J. Umrigar, Phys. Rev. A, {\bf 56}, 290 (1997). 
\item\label{levy3} M. Levy and J.P. Perdew, Phys. Rev. A {\bf 32}, 2010
(1985).
\item\label{kimbal} A.K. Rajagopal, J.C. Kimball and M. Banerjee, 
Phys. Rev. B {\bf 18}, 2339 (1978); J.C. Kimball, Phys. Rev. A 
{\bf 7}, 1648 (1973).
\item\label{matthew} L.M. Fraser {\it et al.}, Phys. Rev. B {\bf 53}, 1814
(1996). 
\item\label{andrew2} A.J. Williamson {\it et al.}, Phys. Rev. B 
{\bf 55}, R4851 (1997). 
\item\label{fulde} R.M. Dreizler and E.K.U. Gross, 
{\em Density Functional Theory, An approach to the Quantum Many-Body
Problem}, (Springer-Verlag, Berlin 1990).
\item\label{hole} J.P. Perdew and Y. Wang, Phys. Rev. B {\bf 46},
12947 (1992).
\item\label{pd3} J.P. Perdew and K. Burke, private communication.
\end{list}
\end{document}